\documentclass[aps,twocolumn,prc,floatfix,showpacs,preprintnumbers,amsmath,amssymb,nofootinbib,groupedaddress]{revtex4-1}
\usepackage{dsfont}
\usepackage{epsfig}
\usepackage{esint}
\usepackage{ulem}
\usepackage[makeroom]{cancel}
\usepackage{wasysym}
\usepackage{color}
\usepackage{graphicx}
\usepackage{dcolumn}    
\usepackage{bigstrut}
\usepackage{amsmath,amsfonts,amsthm,bm}
\usepackage{CJK}
\usepackage[pdfstartview=FitH,CJKbookmarks=true,bookmarksnumbered=true,bookmarksopen=true,colorlinks,pdfborder=001,linkcolor=blue,anchorcolor=blue,citecolor=blue]{hyperref}

\newcommand{\ket}{\rangle}
\newcommand{\bra}{\langle}

\def\be{\begin{equation}}
\def\ee{\end{equation}}
\def\beo{\begin{equation}}
\def\eeo{\end{equation}}
\def\bea{\begin{eqnarray}}
\def\eea{\end{eqnarray}}
\def\bse{\begin{subequations}}
\def\ese{\end{subequations}}

\begin{document}

\title{ Double charge-exchange phonon states}

\author{X. Roca-Maza$^{1,2}$}
\email{xavier.roca.maza@mi.infn.it}
\author{H. Sagawa$^{3,4}$}
\email{sagawa@ribf.riken.jp}
\author{G. Col\`o$^{1,2}$}
\email{colo@mi.infn.it}

\affiliation{$^{1}$Dipartimento di Fisica ``Aldo Pontremoli'', Universit\`a degli Studi di Milano, 20133 Milano, Italy\\
             $^{2}$INFN,  Sezione di Milano, 20133 Milano, Italy\\
             $^{3}$RIKEN Nishina Center, Wako 351-0198, Japan\\   
             $^{4}$Center for Mathematics and Physics, University of Aizu, Aizu-Wakamatsu, Fukushima 965-8560, Japan }

\date{\today}

\begin{abstract}
  We study double charge-exchange phonon states in neutron-rich nuclei, in particular the double isobaric analog states and the double Gamow-Teller  excitations, induced by the double isospin operator  $\sum_{i,j=1}^At_-(i) t_-(j)$ and spin-isospin operator $\sum_{i,j=1}^A{\bm \sigma}(i) t_-(i){\bm \sigma}(j) t_-(j)$, respectively. We employ quartic commutator relations to evaluate the average energies $E_{\rm DIAS} - 2E_{\rm IAS}$ and $E_{\rm DGTR}-E_{\rm DIAS} - 2(E_{\rm GTR}-E_{\rm IAS})$, and conventional double commutator relations to evaluate the average energies of $E_{\rm GTR}-E_{\rm IAS}$ and $E_{\rm IAS}$. We have found that the corrections due to quartic commutators follow the approximate laws: $E_{\rm DIAS} - 2E_{\rm IAS}\approx \frac{3}{2} A^{-1/3}$ MeV and $E_{\rm DGTR}-E_{\rm DIAS} - 2(E_{\rm GTR}-E_{\rm IAS})\approx 16 A^{-1}$ MeV. While the former is dominated by direct Coulomb effects, since Coulomb exchange cancels out to some extent with isospin symmetry breaking contributions originated form the nuclear strong force, the latter is sensitive to the difference in strength between the spin and spin-isospin channels of the strong interaction. 
\end{abstract}

\pacs{24.30.Cz, 11.55.Hx, 21.10.Sf}

\maketitle
\section{Introduction}
The possibility to induce double charge-exchange (DCX) excitations by means of heavy-ion beams at intermediate energies  \cite{RIKEN, INFN-Catania} has recently fostered the interest on new collective excitations such as double isobaric analog states (DIAS) and double Gamow-Teller giant resonances (DGTR). In the 1980s, DCX reactions were performed by using pion beams, i.e., $(\pi^+, \pi^-)$ and $(\pi^-, \pi^+)$ reactions have been studied. Through these experimental investigations, the DIAS, the dipole giant resonance built on the isobaric analog state (IAS) and the double dipole resonance states were identified \cite{kaletka1987,Mord-91,Ward-93,Chomaz-95}. However, the DGTRs were not found in the pion  double charge-exchange spectra.  In the middle of the 1990s, heavy-ion DCX experiments were performed at energies of 76 and 100 MeV/u, with the hope that the DGTR might be observed in the  $^{24}$Mg($^{18}$O, $^{18}$Ne)$^{24}$Ne reaction \cite{MSU-18O}. However, no clear evidence of DGTR was found in this reaction. This is mainly because the  ($^{18}$O, $^{18}$Ne) reaction is a $(2n, 2p)$ type reaction, and even  the single GTR in the $t_+$ channel induced by the $(n,p)$ reaction is weak in $N=Z$ nuclei such as $^{24}$Mg. A research program based on a new reaction, namely ($^{12}$C, $^{12}$Be(0$^+_2$)) has been planned at the RIKEN RIBF facility with high intensity heavy-ion beams at the optimal energy of E$_{\rm lab}$ = 250 MeV/nucleon to excite the spin-isospin response \cite{RIKEN-new}. A big advantage of this reaction  is based on the fact that it is  a $(2p,2n)$ type DCX reaction and one can use  a neutron-rich target to excite DGTR strength. Although many theoretical efforts have been devoted to studies of double $\beta$-decays, DGTR strengths corresponding to the double $\beta$-decays are still too small to be identified in these experiments.  Recently, shell-model calculations were performed to study the DGTR of $^{48}$Ti \cite{Shimizu2018}, 
 and also Ti-isotopes \cite{Auerbach2018}.  At the same time, other DGTR strength distributions have been studied by using the sum rule approach \cite{Vogel,Zamick,Muto,SU2016}, in order to establish a possible unit cross section of DGTR in comparison with the DIAS. Minimally-biased theoretical predictions based on sum rules will  provide a robust and global view of the DGTR, and can be a good guideline for the future experimental studies.

In this paper, we present some formulas to evaluate different combinations of the average excitation energies of the DIAS and DGTR, by using commutator relations  for the double isospin $\sum_{i,j=1}^At_-(i)t_-(j)$ and spin-isospin operator $\sum_{i,j=1}^A{\bm \sigma}(i) t_-(i){\bm \sigma}(j) t_-(j)$. Here ${\bm t}={\bm \tau}/2$,  and ${\bm \sigma}$ and ${\bm \tau}$ denote the Pauli matrices in spin and isospin space, respectively. Specifically, we present formulas to estimate $E_{\rm DIAS}-2E_{\rm IAS}$ from the most relevant Isospin Symmetry Breaking (ISB) terms in the nuclear Hamiltonian and $E_{\rm DGTR}-E_{\rm DIAS} - 2(E_{\rm GTR}-E_{\rm IAS})$ from a simple albeit realistic Hamiltonian including separable residual interactions.

\section{Double Isobaric Analog State}
\label{theo:dias}

\subsection{Average Energy}

The expectation value for the energy of the DIAS is defined as
\begin{equation}
E_{\rm DIAS}\equiv \langle {\rm DIAS}\vert\mathcal{H}\vert{\rm DIAS}\rangle - \langle 0\vert\mathcal{H}\vert 0\rangle \ , 
\label{eq1}
\end{equation}
where $\vert 0\rangle$ represents the ground state and 
\begin{equation}
\vert{\rm DIAS}\rangle \equiv \frac{T_- \vert {\rm IAS}\rangle}{\langle {\rm IAS}\vert T_+T_-\vert{\rm IAS}\rangle^{1/2}}
\label{eq2}
\end{equation}
is the definition of the DIAS state in terms of the IAS that, in turn, can be written as 
\begin{equation}
\vert{\rm IAS}\rangle \equiv \frac{T_- \vert 0\rangle}{\langle 0\vert T_+T_-\vert 0\rangle^{1/2}}\ .
\label{eq3}
\end{equation}
$T_+=\sum_i^A t_+(i)$ and $T_-=\sum_i^A t_-(i)$ are the isospin raising and lowering operators, respectively,  that follow the usual SU(2) algebra;
\be
[T_+,T_-]=2T_z ,  \, \, \, \, [T_z,T_{\pm}]=\pm T_{\pm}, 
\ee where $T_z=\sum_i^A t_z(i)$ and $t_z$ has eigenvalues $-1/2$ for protons and $1/2$ for neutrons. This formulation is general since no assumption is needed for $\mathcal{H}$. 

Starting from Eq.~(\ref{eq1}) and the definitions of the DIAS and IAS previously given, one may write the excitation energy of the DIAS as
\begin{equation}
E_{\rm DIAS} = \frac{\langle 0\vert[T_+^2,[\mathcal{H},T_-^2]] \vert 0\rangle}{\langle 0\vert T_+^2T_-^2\vert 0\rangle} \ , 
\label{eq4}
\end{equation}
assuming that the ground state has good isospin, namely that there is no isospin mixing and $T_+\vert 0\rangle = 0$ (see Appendix \ref{im} for a discussion on isospin mixing effects on IAS and DIAS energies). One can elaborate on the previous equation, and write for the denominator,  
\bea
\langle 0\vert T_+^2T_-^2\vert 0\rangle &=&
\langle 0\vert 4T_z(2T_z-1)\vert 0\rangle\nonumber\\&=&2(N-Z)(N-Z-1),  
\label{eq5}
\eea
whereas the numerator can be expressed as 
\bea
&&\langle 0\vert[T_+^2,[\mathcal{H},T_-^2]] \vert 0\rangle=    \nonumber  \\
&&~~~~~~~~~~= \langle 0\vert 4(2T_z-1) [T_+,[\mathcal{H},T_-]] \vert 0\rangle  \nonumber  \\ 
&&~~~~~~~~~~+ \langle 0\vert[T_+,[T_+,[[\mathcal{H},T_-],T_-]]] 
\vert 0\rangle. 
\label{eq6}
\eea
Remembering that the $E_{\rm IAS}$ is, within the same approximation (i.e. 
no isospin mixing in the ground state),
\begin{equation}
E_{\rm IAS} = \langle {\rm IAS}\vert\mathcal{H}\vert{\rm IAS}\rangle - \langle 0\vert\mathcal{H}\vert 0\rangle = \frac{\langle 0\vert[T_+,[\mathcal{H},T_-]] \vert 0\rangle}{\langle 0\vert T_+T_-\vert 0\rangle}, 
\label{eq7}
\end{equation}
one can eventually write
\begin{equation}
E_{\rm DIAS} = 2E_{\rm IAS} + \frac{\langle 0\vert[T_+,[T_+,[[\mathcal{H},T_-],T_-]]] \vert 0\rangle}{2(N-Z)(N-Z-1)}.
\label{eq8}
\end{equation}

The second term at the right hand side could be different from zero only for ISB terms in $\mathcal{H}$, in a similar manner as they only contribute to $E_{\rm IAS}$ [cf. Eq.~(\ref{eq7})]. In other words, the IAS and DIAS energies are a special filter for the terms in the Hamiltonian that break isospin symmetry (Coulomb and the small contributions from the strong force), while the isospin-conserving part of $\mathcal{H}$ does not contribute and we do not need to specify its form.

The simplest ISB two-body potentials in the nuclear Hamiltonian are proportional to $t_z(1)+t_z(2)$ [Charge Symmetry Breaking (CSB) force]  and to $t_z(1)t_z(2)$ [Charge Independence Breaking (CIB) force]. In the CSB case, the quartic commutator in the numerator of the second term at the right hand side of Eq.~(\ref{eq8}) will give
\begin{eqnarray}
  &&\langle 0\vert[t_+(a),[t_+(b),[[t_z(1)+t_z(2),t_-(c)],t_-(d)]]] \vert 0\rangle  \nonumber \\
  &&~~~~=\langle 0\vert[t_+(a),[t_+(b),[-t_-(1)-t_-(2),t_-(d)]]] \vert 0\rangle = 0.\nonumber\\ 
\label{eq9}
\end{eqnarray}
In other terms, no contribution survives from CSB forces. Note that CSB terms do contribute to the double commutator in Eq.~(\ref{eq7}) as shown in Ref.~\cite{roca-maza2018}. The CIB terms of the specific type $t_z(1)t_z(2)$ will lead after some algebra to 
\bea
&&\langle 0\vert[t_+(a),[t_+(b),[[t_z(1)t_z(2),t_-(c)],t_-(d)]]] \vert 0\rangle   \nonumber  \\ 
&&~~~~=4\langle 0\vert \left\{4t_z(1)t_z(2)-\left[t_+(1)t_-(2)+t_-(1)t_+(2)\right]\right\}\vert 0\rangle\nonumber\\
&&~~~~=8\langle 0\vert \left[3t_z(1)t_z(2)-{\bm t}(1)\cdot{\bm t}(2)\right]\vert 0\rangle \ . 
\label{eq10}
\eea
Hence, CIB interactions will contribute to the quartic commutator in Eq. (\ref{eq8}). In addition to that, we note that other types of  CIB interactions are given  by the operators $T_{ij}\equiv {\bm t}(i)\cdot{\bm t}(j)-3 t_z(i) t_z(j)$, which is a tensor in isospin space, and also by $s_i s_j T_{ij}$, where $s_i=\frac{1}{2}\sigma_i$, and by $S_{ij}T_{ij}$, where $S_{ij}$ is the  tensor operator analogous to $T_{ij}$ but in spin-space. These three operator dependences are implemented in realistic nucleon-nucleon potentials (cf. Ref.~\cite{av18}). However, any of the  CIB terms with $T_{ij}$, if implemented in connection with a zero-range interaction treated at the Hartree-Fock level, will give no contribution to the equation of state (EoS) of symmetric nuclear matter. This would be a drawback since finite-range ISB interactions as those of Ref.~\cite{av18} are known to contribute to the EoS of symmetric nuclear matter \cite{muther1999}. On the other hand, the CIB interaction with $t_z(1)t_z(2)$ dependence gives a finite contribution to the nuclear matter EoS even in the zero-range case  \cite{roca-maza2018}.  This is the reason why we adopt a CIB zero-range interaction of the form shown below [cf. Eq. (\ref{VCIB})], which effectively takes into account those effects into the EoS.

\subsection{The Coulomb contribution}
It is well known that the largest ISB term in the nuclear Hamiltionian is due to the Coulomb interaction, 
\begin{equation}
V_C(\vec{r}_1,\vec{r}_2) = \frac{e^2}{\vert \vec{r}_1-\vec{r}_2 \vert}
[\frac{1}{2}-t_z(1)][\frac{1}{2}-t_z(2)] \ . 
\label{eq11}
\end{equation}

\subsubsection{Direct term}\label{subsec:coul_dir}
\begin{figure}[t!]
\centering
\includegraphics[width=\linewidth,clip=true]{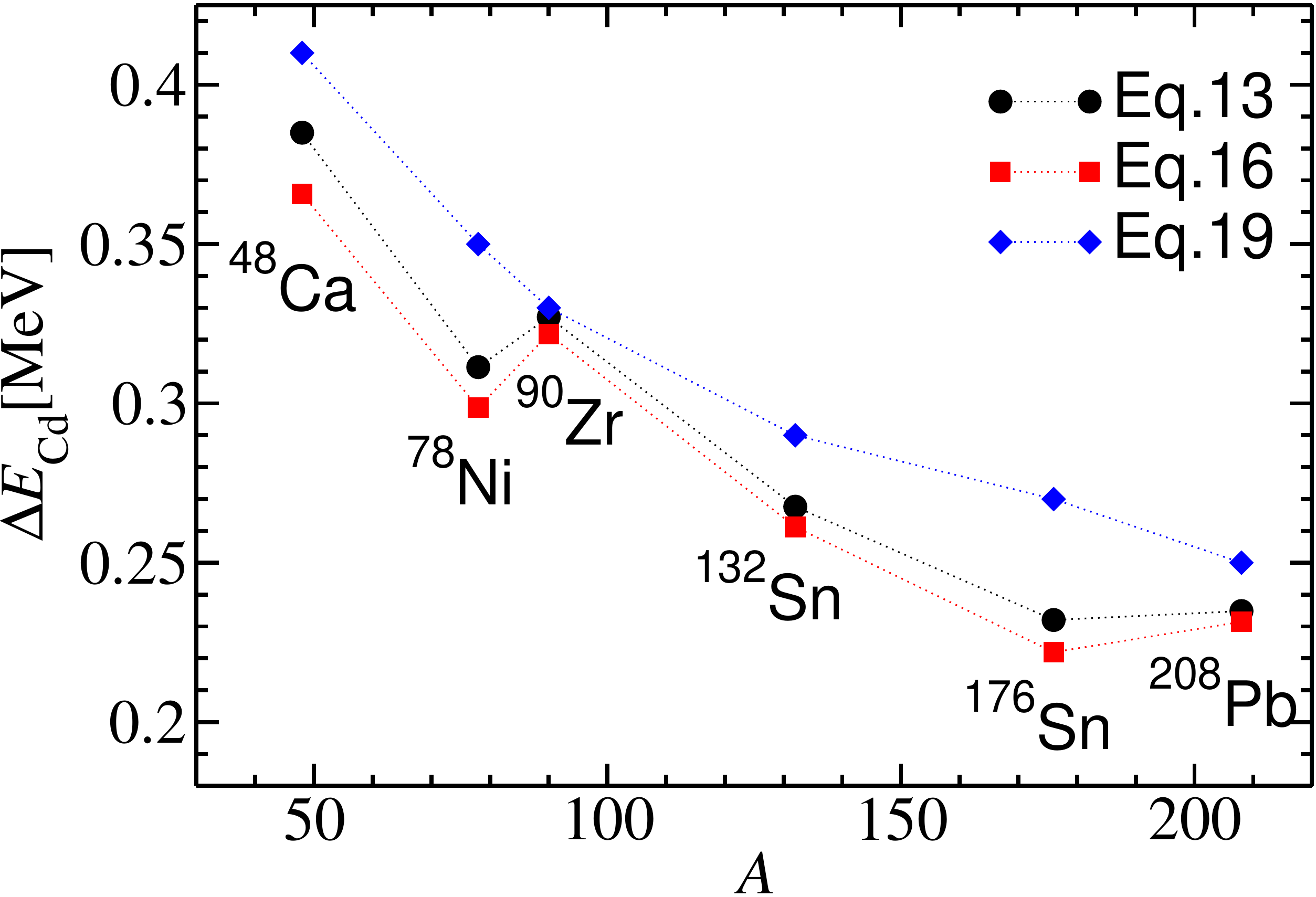} 
\caption{Contribution of the Coulomb direct term to $E_{\rm DIAS}-2E_{\rm IAS}$ as calculated from Eqs. (\ref{eq12}), (\ref{eq13}) and (\ref{eq16}). The SAMi interaction is employed \cite{roca-maza12b}.}
\label{fig1}
\end{figure}

The only non-zero contribution of the Coulomb direct term to the quartic commutator in Eq.~(\ref{eq8}) has the same structure as Eq. (\ref{eq10}). Thus, assuming an independent particle model, we can evaluate the Coulomb direct contribution $\Delta E_{\rm Cd}$ to $E_{\rm DIAS}-2E_{\rm IAS}$ from Eq.~(\ref{eq8}) as follows 
\bea
\Delta E_{\rm Cd}&=&E_{\rm DIAS}-2E_{\rm IAS} \nonumber  \\ 
&=&\frac{\iint d\vec{r}_1d\vec{r}_2\frac{e^2 [\rho_n(\vec{r}_1)-\rho_p(\vec{r}_1)]}{\vert \vec{r}_1-\vec{r}_2 \vert}[\rho_n(\vec{r}_2)-\rho_p(\vec{r}_2)]}{(N-Z)(N-Z-1)} \ . \nonumber \\
\label{eq12}
\eea
Based on the latter result, one can build a very simple and qualitative model to evaluate the right hand side of Eq.~(\ref{eq12}). The model is as follows. We assume that the neutron and proton distributions can be well approximated by a sharp sphere of radius $R_n$ and $R_p$, respectively. The integrals in the coordinate of particle 1 are
\bea
\int d\vec{r}_1\frac{e^2\rho_n(\vec{r}_1)}{|
\vec{r}_1 -\vec{r}|
}=   \left \{
   \begin{array}{ll}
   \frac{Ne^2}{2R_n}    \left(3-\frac{r^2}{R_n^2}\right),    & \quad \text{for } r< R_n\\
    \frac{Ne^2}{r}, & \quad \text{for } r>R_n  
  \end{array}   \right.
\eea
and
\bea
\int d\vec{r}_1\frac{e^2\rho_p(\vec{r}_1)}{\vert \vec{r}_1 -\vec{r} \vert}=  \left \{
  \begin{array}{ll}
    \frac{Ze^2}{2R_p}\left(3-\frac{r^2}{R_p^2}\right),       & \quad \text{for } r< R_p\\
    \frac{Ze^2}{r}. & \quad \text{for } r>R_p 
  \end{array}    \right.
\eea
Therefore, defining $R_n\equiv R_p+\Delta R_{np}$ one can easily find
\bea
\Delta E_{\rm Cd}\approx \sqrt{\frac{3}{5}}\frac{6}{5}\frac{e^2}{\langle r_p^2\rangle^{1/2}}\frac{N-Z}{N-Z-1}\left(1-\frac{N}{N-Z}\frac{\Delta r_{np}}{\langle r_p^2\rangle^{1/2}}\right)\ , \nonumber\\ 
\label{eq13}
\eea
where $\sqrt{5/3}\Delta R_{np}=\Delta r_{np}\equiv \langle r_n^2\rangle^{1/2}-\langle r_p^2\rangle^{1/2}$, and $\langle r_p^2\rangle^{1/2}=\sqrt{\frac{3}{5}}R_p$ within our model. In Fig.~\ref{fig1} and Table~\ref{tab1} some results for the energy difference  $E_{\rm DIAS}-2E_{\rm IAS}$ are given as examples. Specifically, these have been extracted from Eqs.~(\ref{eq12}) and (\ref{eq13}) in the case of some double magic, neutron-rich nuclei. The Skyrme functional SAMi \cite{roca-maza12b} has been employed to calculate densities and corresponding radii. In Table~\ref{tab1} we also show experimental IAS energies and compare them to the energies calculated by means 
of Eq.~(\ref{eq7}) and by taking into account only the Coulomb direct term (in practice, using Eq. (5) of Ref. \cite{PhysRevLett.23.484}).

\begin{table*}[!th]
\caption{Experimental and calculated IAS energies and  calculated DIAS correction energies. Eq.~(\ref{eq7}) is used to calculate the IAS energy by taking into account only the direct Coulomb term. The correction energy $E_{\rm DIAS}-2E_{\rm IAS}$, or more precisely the part due to the Coulomb direct term $\Delta E_{\rm Cd}$, is calculated with the help of Eqs.~(\ref{eq12}),~(\ref{eq13}) and~(\ref{eq16}). The SAMi \cite{roca-maza12b} functional is employed. The experimental  value of the DIAS energy measured from the ground state of $^{48}$Ca is taken from Ref.~\cite{kaletka1987}.   Values are given in MeV.}
\label{tab1}
\centering
\begin{tabular}{l|cc|ccc|cc}
\hline\hline
Nucleus & \multicolumn{2}{c|}{$E_{\rm IAS}$}& \multicolumn{3}{c|}{$\Delta E_{\rm Cd}$} &  \multicolumn{2}{c}{$E_{\rm DIAS}$(exp.)}
 \\  \hline 
               & Exp.   &   Coul. dir.
                 &  Eq.~(\ref{eq12})&Eq.~(\ref{eq13})&Eq.~(\ref{eq16})    &2$E_{\rm IAS}$(Exp)+$\Delta E_{\rm Cd}$(Eq.~(\ref{eq12}))&\\
\hline  
${}^{48}$Ca  &7.182(8)   & 7.20  & 0.385 & 0.366 & 0.413 & 14.749  (14.67) &\\
${}^{78}$Ni  &           & 8.87  & 0.311 & 0.299 & 0.351 & &\\
${}^{90}$Zr  &11.901(12) & 12.23 & 0.327 & 0.322 & 0.335 &24.129 &\\
${}^{132}$Sn &           & 13.60 & 0.268 & 0.261 & 0.295 & &\\
${}^{176}$Sn &           & 12.25 & 0.232 & 0.222 & 0.268 & &\\
${}^{208}$Pb &18.826(10) & 19.45 & 0.235 & 0.231 & 0.253 &37.887 & \\
\hline\hline
\end{tabular}
\end{table*}

In an even simpler manner, within the Liquid Drop Model, $E_{\rm IAS}$ can be estimated as the Coulomb energy difference between the mother ($m$ with $Z_m=Z$) and daughter ($d$ with $Z_d=Z+1$) nucleus, 
\begin{equation}
E_{\rm IAS} = \frac{3}{5}\frac{e^2}{R_p^d}Z_d(Z_d-1) - \frac{3}{5}\frac{e^2}{R_p^m}Z_m(Z_m-1) = \frac{6}{5}\frac{e^2Z}{R_p},
\label{eq14}
\end{equation}
if we assume that $R_{\rm ch}\approx R_p$ and that $R_p^d\approx R_p^m\equiv R_p$.

By using the same model and approximations
\begin{equation}
E_{\rm DIAS} = \frac{3}{5}\frac{e^2}{R_p^d}(Z+2)(Z+1) - \frac{3}{5}\frac{e^2}{R_p^m}Z(Z-1) = \frac{6}{5}\frac{e^2(2Z+1)}{R_p};
\label{eq15}
\end{equation}
therefore, a more crude estimate for the DIAS correction energy reads
\begin{equation}
\Delta E_{\rm Cd}\approx \frac{6}{5}\frac{e^2}{R_p}=\frac{6}{5}\frac{e^2}{r_0}A^{-1/3}\approx\frac{3}{2}A^{-1/3} ~~{\rm MeV}.
\label{eq16}
\end{equation}
By inspecting Eqs.~(\ref{eq13}) and (\ref{eq16}), one can see that Eq.~(\ref{eq13}) essentially corrects Eq.~(\ref{eq16}) by means of the factor within parenthesis that depends on the neutron skin thickness, $\Delta r_{np}$, and that gives the correct trend predicted by Eq. (\ref{eq12}) as compared to the smooth prediction given in Eq.~(\ref{eq16}). The three calculations shown in Fig.~\ref{fig1} coincide very well for ${}^{90}$Zr, that is the nucleus shown in the figure with the smallest isospin asymmetry.

Other contributions to $E_{\rm DIAS}-2E_{\rm IAS}$ exist. From our recent study \cite{roca-maza2018} and previous experience \cite{auerbach1972} on the IAS, other relevant terms are the Coulomb exchange and genuine ISB terms from the nuclear strong force that, specifically, could only come from CIB type forces as previously discussed [cf. Eqs. (\ref{eq9}) and (\ref{eq10})].

\subsubsection{Exchange term}
\begin{figure}[t!]
\centering
\centering
\includegraphics[width=\linewidth,
clip=true]{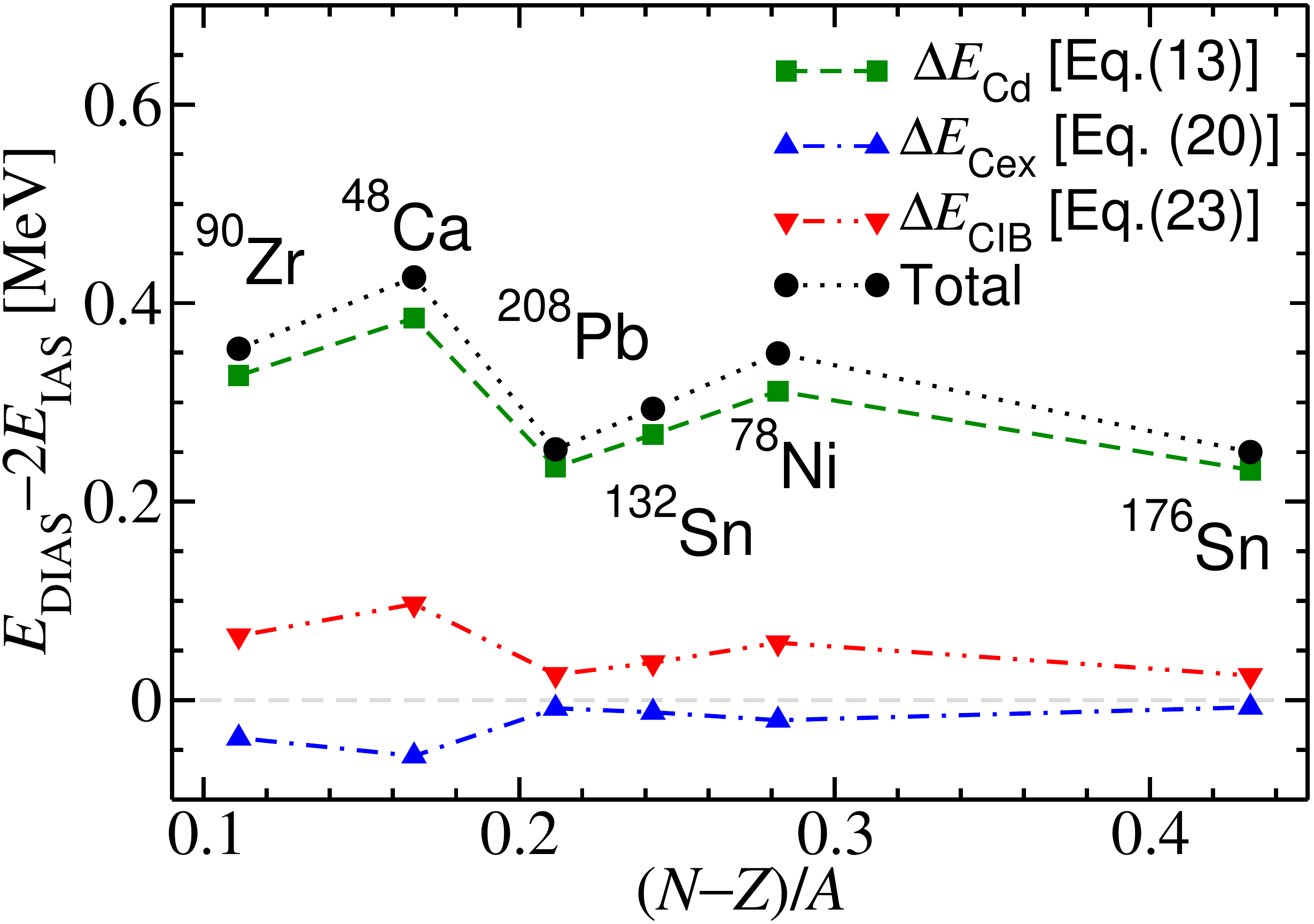} 
\caption{Contribution of the Coulomb direct and exchange terms to $E_{\rm DIAS}-2E_{\rm IAS}$, 
in different neutron-rich nuclei, 
as predicted by SAMi \cite{roca-maza12b}; CIB contribution to $E_{\rm DIAS}-2E_{\rm IAS}$ from Eq. (\ref{e-dias-cib}) as predicted by the CIB term of SAMi-ISB \cite{roca-maza2018}.} 
\label{fig2}
\end{figure}

In what follows, we estimate the Coulomb exchange term. The energy contribution of this term $\Delta E_{\rm Cex}$ to the quartic commutator in Eq. (\ref{eq8}) within an independent particle model reads
\begin{widetext}
\begin{eqnarray}
  \Delta E_{\rm Cex} &=& -4\sum_{ij}\frac{\iint d\vec{r}_1d\vec{r}_2\frac{e^2 t_z(i)t_z(j)}{\vert \vec{r}_1-\vec{r}_2 \vert}\left[\phi_i^*(\vec{r}_1)\phi_j(\vec{r}_1)\phi_j^*(\vec{r}_2)\phi_i(\vec{r}_2)\right]}{(N-Z)(N-Z-1)} \nonumber\\
&&~+\sum_{ij}\frac{\iint d\vec{r}_1d\vec{r}_2\frac{e^2 \left[t_+(i)t_-(j)+t_-(i)t_+(j)\right]}{\vert \vec{r}_1-\vec{r}_2 \vert}\left[\phi_i^*(\vec{r}_1)\phi_j(\vec{r}_1)\phi_j^*(\vec{r}_2)\phi_i(\vec{r}_2)\right]}{(N-Z)(N-Z-1)} \ ,
\label{cex}
\end{eqnarray}
\end{widetext}
where proton and neutron single-particle wave functions contribute. Within the Local Density Approximation (LDA), the contribution to the quartic commutator in Eq. (\ref{eq8}) due to the Coulomb exchange estimated in Eq. (\ref{cex}) can be written as
\begin{widetext}
\bea
\Delta E_{\rm Cex}^{\rm LDA} &=& -\frac{3}{2}\left(\frac{3}{\pi}\right)^{1/3}\frac{e^2}{(N-Z)(N-Z-1)}\int d\vec{r}\Bigg\{\left(\rho_n(\vec{r})-\rho_p(\vec{r})\right)\left(\rho_n(\vec{r})^{1/3}-\rho_p(\vec{r})^{1/3}\right)\nonumber\\
&&\hspace{4cm}-\frac{1}{2}\left(\rho_n(\vec{r})^{2/3}-\rho_p(\vec{r})^{2/3}\right)^2\ln\left(\frac{\rho_n(\vec{r})^{1/3}-\rho_p(\vec{r})^{1/3}}{\rho_n(\vec{r})^{1/3}+\rho_p(\vec{r})^{1/3}}\right)\Bigg\} \ ,
\label{ces}
\eea
\end{widetext}
where none of the terms can be neglected. The contribution of the correction in Eqs. (\ref{cex}) or (\ref{ces}) to $E_{\rm DIAS}-2E_{\rm IAS}$ [Eq. (\ref{eq8})]  is negligible when compared to the Coulomb direct one in Eq. (\ref{eq12}). Some numerical results from Eqs. (\ref{cex}) and (\ref{ces}) based on the SAMi functional are shown in Table~\ref{tab2} and Fig.~\ref{fig2}. 

\begin{table}[b!]
  \caption{Exact (\ref{cex}) and LDA (\ref{ces}) correction to Eq. (\ref{eq8}) due to the Coulomb exchange, as predicted by SAMi \cite{roca-maza12b}; CIB correction to Eq. (\ref{eq8}), from Eq. (\ref{e-dias-cib}) and from the simple model of Eq. (\ref{e-dias-cib-toy}), as predicted by SAMi with $u_0$ and $z_0$ values from SAMi-ISB \cite{roca-maza2018}. All results are for the same nuclei shown in Table \ref{tab1}. Units are MeV.} 
\label{tab2}
\centering
\begin{tabular}{l|cc|cc}
\hline\hline
Nucleus & \multicolumn{2}{c|}{$\Delta E_{\rm Cex}$}&\multicolumn{2}{c}{$\Delta E_{\rm CIB}$}\\   
    & exact   &   LDA & HF & Toy     \\
    &Eq.(\ref{cex})&Eq.(\ref{ces})&Eq.(\ref{e-dias-cib})&Eq.(\ref{e-dias-cib-toy}) \\
\hline
\hline  
${}^{48}$Ca  &-0.056 &-0.033 & 0.097&0.084   \\
${}^{78}$Ni  &-0.020 &-0.017 & 0.058&0.053   \\
${}^{90}$Zr  &-0.038 &-0.020 & 0.065&0.068   \\
${}^{132}$Sn &-0.012 &-0.010 & 0.038&0.039   \\
${}^{176}$Sn &-0.007 &-0.006 & 0.025&0.022   \\
${}^{208}$Pb &-0.008 &-0.007 & 0.026&0.028   \\
\hline\hline
\end{tabular}
\end{table}

\subsection{ISB from the nuclear strong interaction}
As previously discussed, only CIB terms will contribute to the quartic commutator in Eq. (\ref{eq8}). To evaluate their effects, we adopt the recently proposed interaction SAMi-ISB \cite{roca-maza2018}, that has the form
\be
V_{\rm CIB} (\vec{r}_1,\vec{r}_2)=\frac{1}{2}\tau_z(1)\tau_z(2)u_0(1+z_0P_\sigma)\delta(\vec{r}_1-\vec{r}_2) \ , 
\label{VCIB}
\ee
with the parameter $z_0$ fixed to $-1$ and $u_0=25.8$ MeV fm$^3$, fitted to reproduce ISB effects in symmetric nuclear matter as calculated using the Brueckner-Hartree-Fock approach \cite{muther1999} and the realistic nucleon-nucleon interaction AV18 \cite{av18}. In Eq. (\ref{VCIB}) we have introduced the spin-exchange operator $P_{\sigma}$. The energy contribution $\Delta E_{\rm CIB}$ to the quartic commutator in Eq. (\ref{eq8}), from the interaction in Eq. (\ref{VCIB}), within the independent particle model reads
\begin{equation}
\Delta E_{\rm CIB}=\frac{u_0(1-z_0)}{(N-Z)(N-Z-1)}\int d\vec{r}\left(\rho_n-\rho_p\right)^2 \ .
\label{e-dias-cib}
\end{equation}
This result contains both direct plus exchange contributions. Within the simple model previously introduced to estimate the Coulomb direct term, the latter expression can be estimated as 
\begin{equation}
\Delta E_{\rm CIB}\approx\frac{u_0(1-z_0)}{N-Z-1}\rho_0\left(\frac{N-Z}{A}-3\frac{N}{A}\frac{\Delta r_{np}}{\langle r_p^2\rangle^{1/2}}\right) \ ,
\label{e-dias-cib-toy}
\end{equation}
where $\rho_0$ is defined as $\rho_0\equiv 3A/(4\pi R^3)$ and $R=\sqrt{5/3}\langle r^2\rangle^{1/2}$. Numerical results based on the SAMi functional are shown in Table \ref{tab2} and displayed in Fig. \ref{fig2}. It is interesting to note that CIB and Coulomb exchange contributions display the same trends (in absolute value) and cancel to some extent giving a constant contribution to $E_{\rm DIAS}-2E_{\rm IAS}$ of about 30 keV. This correlation can be  understood as follows. Only the CIB terms $\tau_z\cdot\tau_z$ of  Coulomb interaction contributes to the quartic commutator [cf. Eqs.~(\ref{eq9}) and (\ref{eq10})], that is,  the operator structure in isospin space of both  contributions $V_C$ and $H_{CIB}$ is identical so that  their trends should be the same except the absolute values.

From Fig. \ref{fig2}, it is clear that the total contribution to $E_{\rm DIAS}-2E_{\rm IAS}$ from the ISB terms discussed here --Coulomb plus CIB -- is at the level of hundreds of keV with a dependence $\approx A^{-1/3}$ [cf. Fig.~\ref{fig1}, Fig.~\ref{fig2} and Eq.~(\ref{eq16})].

\section{Double Gamow-Teller Resonance}
The non-energy weighted sum rule (NEWSR) for the single Gamow-Teller (GT) transitions is well known and proportional to the neutron excess,
\bea \label{Ikeda}
S_--S_+&=&\sum_f|\bra f|O_-^{\rm GT}|0\ket|^2 - \sum_f|\bra f|O_+^{\rm GT}|0\ket|^2 \nonumber \\
&=&\langle 0\vert[O_+^{\rm GT},O_-^{\rm GT}]\vert 0\rangle = N-Z, 
\eea
where the GT transition operators reads
\be \label{ogt}
O_{\pm}^{\rm GT}=\sum_i^A\sigma_z(i) t_{\pm}(i).
\ee
Notice that there is no factor 3 in front of $N-Z$ in Eq. (\ref{Ikeda}) since we do not sum up the three components of the spin operator $\sigma_\alpha(i)$ ($\alpha=x,y,z$) in the definition of Eq. (\ref{ogt}). All the results from the commutators that follow will not change their structure if we sum up these three components, in spherical nuclei, and will simply be multiplied by a factor 3. To simplify the notation, we will drop in what follows the GT label in $O_\pm^{\rm GT}$. GT operators satisfy the property $O_{\pm}^\dag=O_{\mp}$, and the commutation relations $[O_+,O_-]=2T_z$ and $[T_z,O_{\pm}]=\pm O_\pm$, similar to the isospin operators defined in the previous section. The reason is that the spin matrix $\sigma_z$ commutes with the isospin operators. 

The GT NEWSR is model independent and gives a good guidance when performing the single charge-exchange reactions such as the  $(p,n)$ and ($^3$He, t) reactions with the goal to pin down the GTR strength in nuclei (see, for example, the 
review article of Ref. \cite{Sakai-06}).  

We define the mean energy of the DGTR with respect to the ground state energy, in analogy to the DIAS case, as
 \bea \label{eave}
E_{{\rm DGTR}}&\equiv& \ \langle {\rm DGTR}\vert \mathcal{H}\vert {\rm DGTR}\rangle - \langle 0\vert \mathcal{H}\vert 0\rangle \ ,
\eea 
where the DGTR state is defined as
\begin{equation}
  \vert {\rm DGTR}\rangle\equiv \frac{O_-\vert {\rm GT}\rangle}{\langle {\rm GT}\vert O_+O_-\vert {\rm GT}\rangle^{1/2}}
\end{equation}
and the single GT state as
\begin{equation}
  \vert {\rm GT}\rangle\equiv \frac{O_-\vert 0\rangle}{\langle 0\vert O_+O_-\vert 0\rangle^{1/2}} \ .
\end{equation}
Assuming the parent state $\vert 0\rangle$ has good isospin, that is $T_{+}|0 \rangle=0$, one can write the average excitation energy (\ref{eave}) in a convenient commutator form,
\be \label{eave2}
E_{{\rm DGTR}}=  \frac { \langle  0|[O_{+}^2, [\mathcal{H}, O_{-}^2]]|0 \rangle}{ \langle  0|O_{+}^2 O_{-}^2|0 \rangle} \ .
\ee

The numerator of Eq. (\ref{eave2}) can be expressed as,
\bea  \label{DDCT1}
&&\langle  0|[O_{+}^2,  [\mathcal{H}, O_{-}^2]]|0 \rangle 
=\langle  0|[O_{+},[O_{+}, [[H, O_{-}],O_{-}]]]|0 \rangle  \nonumber \\
&& ~~~~~~~~~+\langle  0|4(2T_z-1)[O_{+}, [H, O_{-}]]|0 \rangle \ ,  
\eea
or equivalently, 
\bea  \label{DDCT2}  
&&\langle  0|[O_{+}^2,  [\mathcal{H}, O_{-}^2]]|0 \rangle 
=\langle  0|[O_{+},[[O_{+}, [H, O_{-}]],O_{-}]]|0 \rangle  \nonumber \\
&& ~~~~~~~~~+\langle  0|2(4T_z-1)[O_{+}, [H, O_{-}]]|0 \rangle \ .
\eea
Note that the result in Eq. (\ref{DDCT1}) has the same structure as that in Eq. (\ref{eq6}), while the result in Eq. (\ref{DDCT2}) differs from it. We introduce Eq. (\ref{DDCT2}) for convenience, as it will be clear below (Appendix \ref{app3}). 

The denominator of Eq. (\ref{eave2}), assuming a parent state with good isospin ($T_+\vert 0\rangle=0$), is given by
\bea \label{E-deno}
\langle  0|O_{+}^2O_{-}^2|0 \rangle &=& \langle  0\vert 4T_z\left(2T_z-1\right)\vert 0 \rangle \nonumber\\
&=& 2(N-Z)(N-Z-1).
\eea
Hence, we can write the energy of the DGTR by using Eqs. (\ref{DDCT1}) and (\ref{E-deno}) as
\be \label{eave3}
E_{\rm DGTR} = 2 E_{\rm GT} + \frac{\langle  0|[O_{+},[O_{+}, [[H, O_{-}],O_{-}]]]|0 \rangle}{2(N-Z)(N-Z-1)}\ ,
\ee
or, equivalently, by using Eqs. (\ref{DDCT2}) and (\ref{E-deno}) as 
\bea
E_{\rm DGTR} &=& \frac{1}{2}\left(\frac{N-Z}{N-Z-1}+1\right)E_{\rm GTR} \nonumber \\
           &+& \frac{\langle  0|[O_{+},[[O_{+}, [H, O_{-}]],O_{-}]]|0 \rangle}{2(N-Z)(N-Z-1)}\ .
\label{eave31}
\eea

In order to evaluate the different quartic and double commutators, we assume the following general form for the Hamiltonian,
\be \label{Hamil}
\mathcal{H}=\mathcal{H}_0+V+V_C+V_{\rm ISB} \ ,
\ee
where $V$ is the spin- and isospin-dependent interaction, $V_C$ is the Coulomb interaction and $V_{\rm ISB}$ is an ISB effective interaction originated from the nuclear strong force, as the one we have used above from Ref. \cite{roca-maza2018}. $\mathcal{H}_0$ is the spin and isospin independent part of the Hamiltonian. From Eqs. (\ref{eave2}) and (\ref{Hamil}), we can derive the relation between the DGTR and the DIAS, 
\be \label{Hamil2}
E_{\rm DGTR}-E_{\rm DIAS}=\frac{ \langle  0|[O_{+}^2, [V, O_{-}^2]]|0 \rangle}{2(N-Z)(N-Z-1)} \ ,
\ee
since
\be \label{DDC1}
 [ O_{+}^2,  [V_C+V_{\rm ISB},  O_{-}^2]]=[{\rm T}_+^2,  [V_C+V_{\rm ISB}, {\rm T}_-^2]] \ .
 \ee
Introducing the DIAS energy is convenient here, as it allows one to isolate the effect of the spin- and isospin-dependent interaction $V$ in the quantity $E_{\rm DGTR}-E_{\rm DIAS}$, exactly in the same way as in the difference of its single charge-exchange counterpart $E_{\rm GT}-E_{\rm IAS}$.
Using Eqs. (\ref{DDCT1}) and (\ref{eave3}), one can rewrite Eq. (\ref{Hamil2}) as
\bea \label{Hamil31}
E_{\rm DGTR}&-&E_{\rm DIAS}-2(E_{\rm GT}-E_{\rm IAS})=\nonumber \\ &=&\frac{\langle  0|[ O_{+},[ O_{+}, [[V,  O_{-}], O_{-}]]]|0 \rangle}{2(N-Z)(N-Z-1)}\ ,
\eea
since 
\bea
&&[ O_{+},[ O_{+}, [[V_C+V_{\rm ISB},  O_{-}], O_{-}]]]=\nonumber\\
&&~~~~ [\hat{T}_{+},[\hat{T}_{+}, [[V_C+V_{\rm ISB}, \hat{T}_{-}],\hat{T}_{-}]] ] , 
\eea
or, equivalently, using Eqs. (\ref{DDCT2}), (\ref{eave3}) and (\ref{Hamil2}), 
\bea
E_{\rm DGTR}-E_{\rm DIAS}&=& \left(1+\frac{N-Z}{N-Z-1}\right)(E_{\rm GTR}-E_{\rm IAS})\nonumber \\ &+& \frac{\langle  0\vert[ O_{+},[[ O_{+}, [V,  O_{-}]], O_{-}]]\vert 0 \rangle}{2(N-Z)(N-Z-1)} \ .
\label{Hamil32}
\eea

A collective state could be represented as a coherent particle-hole superposition induced by a one body, oscillating and self-sustaining, average field, proportional to ${\bm \sigma}\cdot{\bm \tau}$ operators in the GT case. This is equivalent to expressing the two-body interaction in a separable form \cite{BM1,gaarde1983}. In our case, in order to evaluate the energy difference between $E_{\rm DGTR}$ and $E_{\rm DIAS}$, we adopt the following separable interaction \cite{BM1,Suzuki1,Suzuki2} 
 \begin{eqnarray}  
&& V=\sum_i^{A}\kappa_{ls}{\bm l}(i)\cdot{\bm s}(i) +\frac{1}{2}\frac{\kappa_{\tau}}{A}\sum_{i\neq j}^{A}{\bm \tau}(i)\cdot{\bm \tau}(j)   \nonumber \\
 &&+\frac{1}{2}\frac{\kappa_{\sigma}}{A}\sum_{i\neq j}^{A}{\bm \sigma}(i)\cdot{\bm \sigma}(j)  +\frac{1}{2}\frac{\kappa_{\sigma\tau}}{A}\sum_{i\neq j}^{A}({\bm \sigma}(i)\cdot{\bm \sigma}(j)) ({\bm \tau}(i)\cdot{\bm \tau}(j) )\  , \nonumber\\
   \label{interac}
 \end{eqnarray}
 where 
   $\kappa_{ls}$ is the one-body spin-orbit coupling strength while $\kappa_{\tau}$,  $\kappa_{\sigma}$ and  $\kappa_{\sigma\tau}$ are the coupling strengths of the residual two-body interactions in the isospin, spin and spin-isospin channels, respectively.

\begin{table*}[t!]
\caption{Single and double GTR excitation energies referred to the single and double IAS, respectively, for some neutron-rich closed-shell nuclei. $l$ is the angular momentum of the active orbit for the GT excitations and $\Delta \varepsilon_{ls}$ is its  spin-orbit energy splitting. In the next column we provide the spin-orbit contribution $\Delta E_{ls}$ to $(E_{\rm GT}-E_{\rm IAS})$ in Eq. (\ref{GT-E}). $(E_{\rm GT}-E_{\rm IAS})$ from Eq. (\ref{GT-E}) and $E_{\rm DGTR}-E_{\rm DIAS}$ from Eqs. (\ref{EGT-EIAS_2}-\ref{EGT-EIAS_3}) are given in the next columns. In the last column, an estimate of $E_{\rm DGTR}-E_{\rm DIAS}-2(E_{\rm GT}-E_{\rm IAS})$ based on Eq.~(\ref{EGT-EIAS_2}) is also provided. The parameters of the interaction (\ref{interac}) used here are $V_{ls}=34$ MeV and $\kappa_{\sigma\tau}-\kappa_\tau=-4$ MeV. 
}
\label{tab3}
\centering
\begin{tabular}{l|c|cc|c|cc|cc|c}
\hline\hline
Nucl. & $l$ & \multicolumn{2}{c|}{$\Delta \varepsilon_{ls}$}  &$\Delta E_{ls}$ &\multicolumn{2}{c|}{$E_{\rm GT}-E_{\rm IAS}$}& \multicolumn{2}{c|}{$E_{\rm DGTR}-E_{\rm DIAS}$}&$E_{\rm DGTR}-E_{\rm DIAS}$\\ 
 & &  Exp. &Calc. &Eq.(\ref{GT-E}) &  Exp. &Eq.(\ref{GT-E}) & Eq.(\ref{EGT-EIAS_2})&Eq.(\ref{EGT-EIAS_3})&$-2(E_{\rm GT}-E_{\rm IAS})$\\
 & &[MeV]&[MeV]&[MeV]&[MeV]&[MeV]&[MeV]&[MeV]&[MeV]\\
\hline
${}^{48}$Ca    &3&$\sim$ 9& 9.0&5.1 & 3.3&3.8 & 8.0 & 8.0 & 0.38\\      
${}^{90}$Zr    &4&7.54    & 7.6&4.5 & 3.9&3.6 & 7.4 & 7.4 & 0.20\\       
${}^{132}$Sn   &4& --     & 5.9&1.09& -- &0.79& 1.7 & 1.7 & 0.13\\       
              &5& --     & 7.2&1.64&    & &     &     &     \\               
${}^{208}$Pb   &5& 5.60   & 5.3&0.88& 0.4 &0.42& 0.92& 0.92& 0.08\\      
              &6& 5.86   & 6.3&1.23&     & &     &     &     \\               
\hline\hline
\end{tabular}
\end{table*}

The average energy of the GTR minus that of the IAS is expressed as (cf. Appendix \ref{app2})
 \bea \label{GT-E}
E_{\rm GT}&-&E_{\rm IAS}=\frac{ \langle  0|[ O_{+}, [V,  O_-]]|0 \rangle}{(N-Z)} \nonumber \\
&=&-\frac{4}{3}\frac{\kappa_{ls}}{N-Z}\langle  0|\sum_i^{A}{\bm l}(i)\cdot{\bm s}(i)|0 \rangle \nonumber \\
&+&2( \kappa_ {\sigma \tau}- \kappa_ {\tau}) \frac{N-Z}{A}\ .
\eea

In a similar way, the energy difference between DGTR and DIAS (\ref{Hamil2}) is expressed as (cf. Appendix \ref{app3})
 \bea \label{EGT-EIAS_1}
 &&E_{\rm DGTR}-E_{\rm DIAS}-\left(1+\frac{N-Z}{N-Z-1}\right)(E_{\rm GT}-E_{\rm IAS})=\nonumber\\
&&~~~~=\frac{4}{3}\frac{\kappa_{ls}\langle  0|\sum_i^{A}{\bm l}(i)\cdot{\bm s}(i)|0 \rangle}{(N-Z)(N-Z-1)}
\nonumber \\
&&~~~~-6( \kappa_ {\sigma \tau}- \kappa_ {\tau}) \frac{1}{A}\frac{N-Z}{N-Z-1}\ .
\eea
This, after some algebra, can be rewritten as  
 \bea \label{EGT-EIAS_2}
 &&E_{\rm DGTR}-E_{\rm DIAS}-2(E_{\rm GT}-E_{\rm IAS})=\nonumber\\
&&~~~~=-4\frac{\kappa_ {\sigma \tau}- \kappa_ {\tau}}{A}\frac{N-Z}{N-Z-1}\ .
\eea
In turn, 
the latter expression can be also written as follows within our model, 
 \bea \label{EGT-EIAS_3}
 &&E_{\rm DGTR}-E_{\rm DIAS}-2\frac{N-Z-2}{N-Z-1}(E_{\rm GT}-E_{\rm IAS})=\nonumber\\
&&~~~~=-\frac{8}{3}\frac{\kappa_{ls}\langle  0|\sum_i^{A}{\bm l}(i)\cdot{\bm s}(i)|0 \rangle}{(N-Z)(N-Z-1)}\ , 
\eea
and
the advantage of this expression is that it does not explicitly depend on the isospin $\kappa_\tau$ and spin-isospin $\kappa_{\sigma\tau}$ coupling strengths. Hence, if the experimental value of $E_{\rm GT}-E_{\rm IAS}$ is known, one may easily estimate $E_{\rm DGTR}-E_{\rm DIAS}$ based on Eq. (\ref{EGT-EIAS_3}) and on a reasonable single-particle level scheme close to the Fermi surface.  

In order to theoretically estimate, with our simple yet physical model, the value of $E_{\rm DGTR}-E_{\rm DIAS}$, we proceed as follows. For the spin-orbit coupling, which is surface-dominated, we adopt a formula with an $A^{2/3}$ dependence \cite{BM1},  
  \bea
   \kappa_{ls}=-V_{ls}/A^{2/3} \ , 
 \eea
where the coupling $V_{ls}$ has been adjusted to reproduce the experimental values of the spin-orbit splittings $\Delta \varepsilon_{ls}$ of some active orbits for the GT excitations for the nuclei given in Table \ref{tab3}. The optimal value found for the spin-orbit strength parameter is $V_{ls}=34$ MeV, and the corresponding results can be also seen in the same Table. 

In order to fix $\kappa_{\sigma \tau}-\kappa_\tau$ we adopt a similar strategy. Assuming $V_{ls}=34$ MeV, we find the optimal value for $\kappa_{\sigma \tau}-\kappa_{\tau}$ that reproduces the experimental value of $E_{\rm GT}-E_{\rm IAS}$ in ${}^{48}$Ca \cite{yako2009}, ${}^{90}$Zr \cite{wakasa1997, krasznahorkay2001}, ${}^{112-124}$Sn \cite{pham1995} and ${}^{208}$Pb \cite{akimune1995} via Eq. (\ref{GT-E}) (see appendix \ref{app4} for some details). The value found is $\kappa_{\sigma \tau}-\kappa_{\tau}=-4$ MeV in good agreement with previous literature \cite{gaarde1983,osterfeld1992}. In Table \ref{tab3}, we show the contribution of the spin-orbit term $\Delta E_{ls}$ to $E_{\rm GT}-E_{\rm IAS}$ in Eq. (\ref{GT-E}), as well as some results for the single and double GTR when referred to the single and double IAS, respectively, for some doubly-magic nuclei. Specifically, in the 6th and 7th column we provide the experimental $E_{\rm GT}-E_{\rm IAS}$ as well as the estimate from Eq. (\ref{GT-E}) obtained by using the optimal $V_{ls}=34$ MeV and $\kappa_{\sigma\tau}-\kappa_\tau=-4$ MeV values. Next to it, in the 8th and 9th columns we show the corresponding predictions for $E_{\rm DGTR}-E_{\rm DIAS}$ from Eqs. (\ref{EGT-EIAS_2}-\ref{EGT-EIAS_3}). In the last column an estimate of $E_{\rm DGTR}-E_{\rm DIAS}-2(E_{\rm GT}-E_{\rm IAS})$ based on Eq.~(\ref{EGT-EIAS_2}) is also given. The estimated values are of hundreds of keV and account for a few \% correction of the $E_{\rm DGTR}-E_{\rm DIAS}$. Hence, according to our model, this implies that if $E_{\rm DGTR}-E_{\rm DIAS}$ and $E_{\rm GT}-E_{\rm IAS}$ can be determined to a better accuracy than a few \%, DCX measurments of $E_{\rm DGTR}-E_{\rm DIAS}$ will constitute a new way to probe  spin and spin-isospin properties in nuclei.

\section{Summary}

Double GT and IAS average excitation energies have been determined for the first time using double and quartic commutator relations. In order to provide semi-quantitative theoretical estimates, we have adopted two approximations. In the first place, an independent particle picture have been assumed. We have also provided expressions in which, by simplifying further, the neutron and proton distributions have been taken as hard spheres. This simplification has turned out to be very much useful in order to capture the main terms dominating the calculated quantities. 

As a conclusion, within our approach double resonance energies in neutron-rich nuclei are dominated by the same physics of their single counterparts since the main contribution to them is $2E_{\rm IAS}$ and $2E_{\rm GTR}$, respectively. Hence, the effect of two-body Coulomb interaction has a decisive effect on the average energy $E_{\rm DIAS}$, while the spin-orbit and residual isospin and spin-isospin interactions play a big role for the average energy $E_{\rm DGTR}-E_{\rm DIAS}$.  More specifically, we have found that the corrections due to quartic commutators follow the approximate laws: $E_{\rm DIAS} - 2E_{\rm IAS}\approx \frac{3}{2} A^{-1/3}$ MeV (even when the isospin mixing effects are accounted), and $E_{\rm DGTR}-E_{\rm DIAS} - 2(E_{\rm GTR}-E_{\rm IAS})\approx 16 A^{-1}$ MeV. While the former is dominated by Coulomb direct effects since Coulomb exchange cancel out to some extent with isospin symmetry breaking contributions originated form the nuclear strong force, the latter is very sensitive to the difference in strength between the spin and spin-isospin chanels of the strong interaction. Finally, we note that $E_{\rm DGTR}-E_{\rm DIAS} - 2(E_{\rm GTR}-E_{\rm IAS})$ account for a few \% correction ($\lesssim 10$ \%) to the $E_{\rm DGTR}-E_{\rm DIAS}$, implying that if $E_{\rm DGTR}-E_{\rm DIAS}$ and $E_{\rm GT}-E_{\rm IAS}$ can be determined to a better accuracy than a few \%, double charge-exchange measurements of $E_{\rm DGTR}-E_{\rm DIAS}$ will constitute a new promising tool to probe spin and spin-isospin properties in nuclei.

\section*{Acknowledgments}
We would like to thank K. Yako and T. Uesaka for valuable discussions on experimental status of DIAS and DGTR research. This work was supported in part by JSPS KAKENHI  Grant Numbers JP16K05367. Funding from the European Union's Horizon 2020 research and innovation programm under grant agreement No 654002 is also acknowledged.

\appendix

\section{Isospin mixing}
\label{im}
In Sec.~\ref{theo:dias}, we have assumed that there is no isospin mixing in the ground state when calculating the IAS and DIAS energies, that is, $T_+\vert 0\rangle = 0$. This assumption is not exact \cite{auerbach1972}, althought it is a good approximation as we shall see in what follows.

\subsection{Correction to the wave function}
\label{imwf}
The isospin symmetry breaking terms in the Hamiltonian $\mathcal{H}$ can be decomposed into isoscalar, isovector and isotensor. Accordingly, the nuclear ground state $\vert 0\rangle$ can be projected on a basis with good isospin quantum numbers $\vert T T_z\rangle$ as follows, 
\begin{equation}
\vert 0\rangle = \alpha_{T,~T}\vert T,~T\rangle + \alpha_{T+1,~T}\vert T+1,~T\rangle + \alpha_{T+2,~T}\vert T+2,~T\rangle \ .
\label{wf}
\end{equation}
It is expected that the coefficients obey $\alpha_{T,~T}>>\alpha_{T+1,~T}>>\alpha_{T+2,~T}$ (cf. Ref. \cite{auerbach1983} and Eqs. (6.32-6.35) in Ref. \cite{wilkinson}).

 We now  estimate the amount of mixing in the wave function under this hypothesis. The non-energy-weighted sum in the $t_+$ channel reads 
\begin{eqnarray}
  m_0^+&\equiv&\langle 0\vert T_-T_+\vert 0\rangle\nonumber\\
  &=& \langle 0\vert \left(T^2-T_z^2-T_z\right)\vert 0\rangle\nonumber\\
  &=&~~\alpha_{T+1,~T}^2(N-Z+2)\nonumber\\
  &&+2\alpha_{T+2,~T}^2(N-Z+3) \ .
\end{eqnarray}
If $\alpha_{T+2,~T}^2$ can be neglected, then
\begin{equation}
\varepsilon^2 \equiv \alpha_{T+1,~T}^2 = m_0^+/(N-Z+2) \ ,
\end{equation}
(cf. Eq.~(4.29) of Ref.~\cite{auerbach1983}). 

We give numerical results for $\varepsilon^2$ in Table \ref{tab4}. To obtain those values, we evaluate, for different nuclei, the sum rule value of $m_0^+$ by means of the Tamm-Dancoff Approximation (TDA) and the Random Phase Approximation (RPA). All calculations have been based on the SAMi functional.

\begin{table}[!th]
\caption{Estimated values of $\varepsilon^2\equiv\frac{m_0^+}{N-Z+2}$ by using the SAMi interaction. In the second and third columns,  $\varepsilon^2$ has been calculated using the TDA including $V_C$ and $V_C+V_{\rm ISB}$, respectively. In the fourth and fifth columns,  $\varepsilon^2$ has been calculated using the RPA  including $V_C$ and $V_C+V_{\rm ISB}$, respectively. Values are given in \%.} 
\label{tab4}
\centering
\begin{tabular}{l|cc|cc}
\hline\hline
Nucleus & \multicolumn{2}{c|}{TDA}& \multicolumn{2}{c}{RPA} \\  \hline 
        & $V_{C}$&$V_{C}+V_{\rm ISB}$&$V_{C}$&$V_{C}+V_{\rm ISB}$ \\ \hline  
${}^{48}$Ca  & 0.53 & 0.49 & 0.09 & 0.01 \\
${}^{90}$Zr  & 1.05 & 0.97 & 0.44 & 0.31 \\
${}^{132}$Sn & 0.43 & 0.40 & 0.11 & 0.07 \\
${}^{208}$Pb & 0.66 & 0.63 & 0.28 & 0.22 \\
\hline\hline
\end{tabular}
\end{table}

The comparison between TDA and RPA results allows us to quantify the amount of spurious isospin mixing in the ground state wave function used for the calculations in Sec.~\ref{theo:dias}. Specifically, TDA calculations are based 
on the HF ground state that is known to contain both spurious and physical isospin mixing contributions. Self-consistent RPA restores the isospin symmetry, so that the results include only the physical isospin mixing. In Table \ref{tab4}, we  report the effect of the Coulomb interaction $V_C$ on the isospin mixing in the wave function as well as the effect of other ISB terms in the strong interaction $V_{\rm ISB}$, introduced as in Ref.~\cite{roca-maza2018}.

\subsection{Correction to $E_{\rm IAS}$}
\label{imes_ias}

\begin{figure}[t!]
\centering
\includegraphics[width=0.8\linewidth,clip=true]{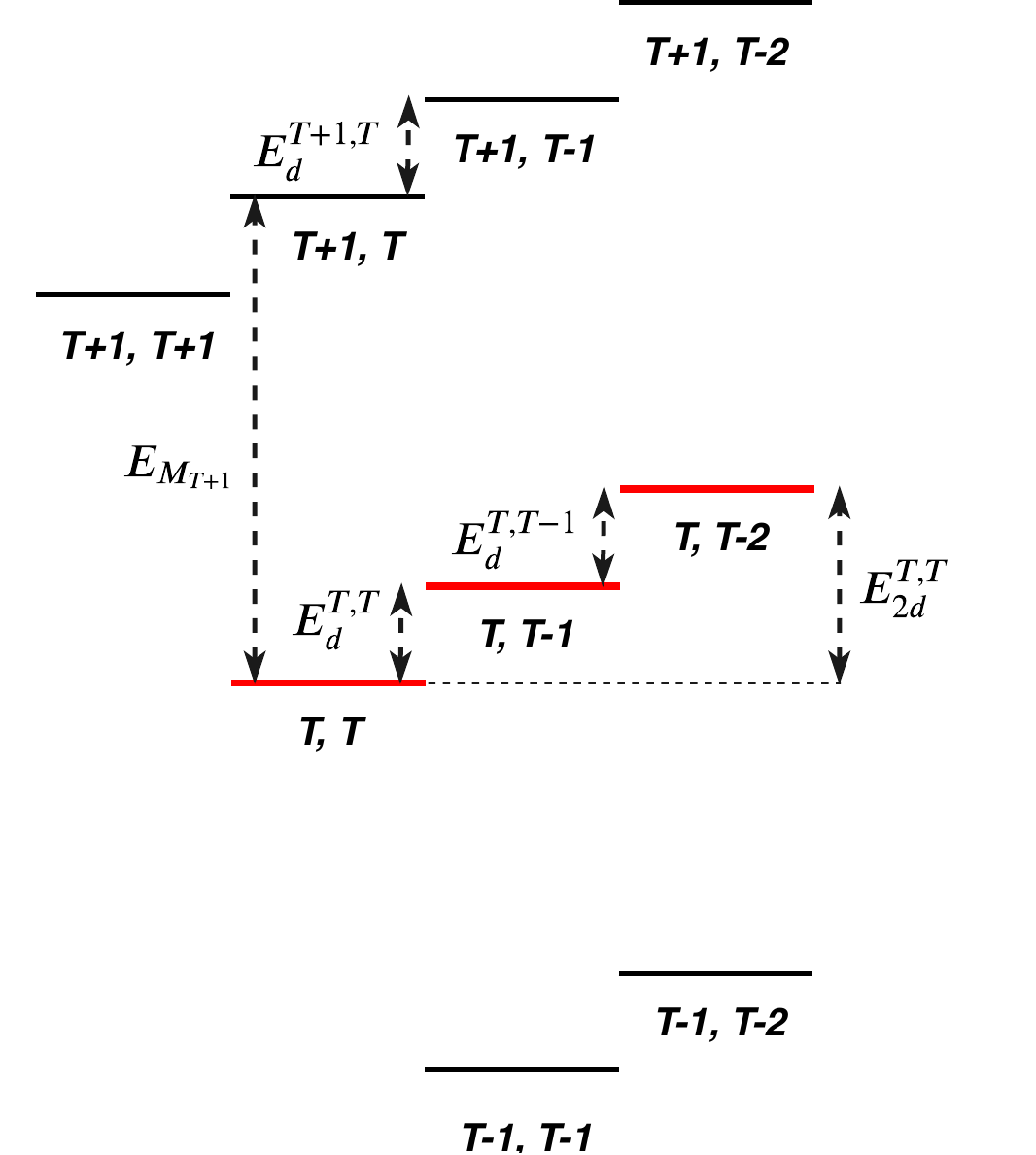} 
\caption{Schematic representation of the energy differences between multiplets of good isospin. Some of the energy differences defined in the text are given explicitly.}
\label{fig3}
\end{figure}

After the  determination of the coefficient $\alpha_{T+1,~T}$,  
  we now  estimate the energy shift due to isospin mixing effects in the IAS as predicted by Eq.~(\ref{eq7}). Assuming that $T_+\vert 0\rangle\neq 0$, one should correct Eq.~(\ref{eq7}) as follows \cite{auerbach1972}, 
\begin{equation}
  E_{\rm IAS} = \frac{\langle 0\vert[T_+,[\mathcal{H},T_-]] \vert 0\rangle}{\langle 0\vert T_+T_-\vert 0\rangle}
+\frac{\langle 0\vert[\mathcal{H},T_-]T_+ \vert 0\rangle}{\langle 0\vert T_+T_-\vert 0\rangle} \ , 
\label{eias:im}
\end{equation}
where the first term in the rhs is the same as in Eq.~(\ref{eq7}) and the second term will contribute only if the ground state wave function $\vert 0\rangle$ contains some isospin mixing effects. 
 The first term in the rhs  contains the effects of  isospin impurities  coming from the numerator and the denominator. 
 The effect of numerator is implicitly included in our numerical results shown in Sec.~\ref{theo:dias} via the employed HF densities,  while those arising from the denominator have been neglected in the same numerical results in Sec.~\ref{theo:dias}. Hence, the isospin mixing correction ($\Delta E_{\rm IAS}$) to our results on the IAS energy reported in Sec.~\ref{theo:dias} (third column in Table \ref{tab1}) can be written as
\begin{eqnarray}
 E_{\rm IAS} &=& \frac{\langle 0\vert[T_+,[\mathcal{H},T_-]] \vert 0\rangle}{N-Z}+\Delta E_{\rm IAS}, 
\end{eqnarray}
where 
\begin{eqnarray}
 \Delta E_{\rm IAS} &=& -\varepsilon^2\frac{N-Z+2}{N-Z}\frac{\langle 0\vert[T_+,[\mathcal{H},T_-]] \vert 0\rangle}{N-Z}\nonumber\\
 &&+\frac{\langle 0\vert[\mathcal{H},T_-]T_+ \vert 0\rangle}{\langle 0\vert T_+T_-\vert 0\rangle} \ ,
\label{eqa12} 
\end{eqnarray}
and where we have considered that
\begin{equation}
\vert 0\rangle = \sqrt{1-\varepsilon^2}\vert T,~T\rangle + \varepsilon\vert T+1,~T\rangle.
\end{equation}
In order to quantitatively estimate Eq.~(\ref{eqa12}), we also assume:
\begin{itemize}
\item[i)] $\mathcal{H}=\mathcal{H}_0+\mathcal{H}_{\rm ISB}$ where $\mathcal{H}_0$ preserves isospin, $\mathcal{H}_0\vert T,~T_z\rangle = E_{T,~T_z}\vert T,~T_z\rangle$, while $\mathcal{H}_{\rm ISB}$ does not. 
\item[ii)] Diagonal contributions of $\mathcal{H}_{\rm ISB}$ are neglected, since they do not mix the  isospin and do not affect markedly our estimates of the isospin mixing correction. That is, we assume that $\langle T+n,~T\vert\mathcal{H}\vert T+n,~T\rangle\approx \langle T+n,~T\vert\mathcal{H}_0\vert T+n,~T\rangle$.
\item[iii)] The other, non-diagonal contributions of $\mathcal{H}_{\rm ISB}$ are approximated using first order perturbation theory. That is, $\varepsilon=\frac{\langle T+1,~T\vert\mathcal{H}_{\rm ISB}\vert T,~T\rangle}{E_{T,~T}-E_{T+1,~T}}$.     
\end{itemize}
Under these assumptions and keeping terms up to $\varepsilon^2$,  we find  
\begin{eqnarray}
  \Delta E_{\rm IAS}&\equiv&-\varepsilon^2\frac{N-Z+2}{N-Z}\Big(E_{T+1,~T}-E_{T,~T}\nonumber\\
&&~~~~~~~~~~~~~~~~~  +E_d^{T,T}-E_d^{T+1,T}\Big) \ ,
\label{eqa10}
\end{eqnarray}
where $E_d^{T,~T}\equiv E_{T,~T-1}-E_{T,~T}$ is the displacement energy of the nucleus with isospin $T$ (see Fig.~\ref{fig3} for a schematic representation). This derivation was previously given in Ref.~\cite{auerbach1983}. The difference in the displacement energies between the nucleus with isospin $T$ and $T+1$ is negligible as compared to the difference $E_{T+1,~T}-E_{T,~T}=E_{M_{T+1}}$, that is nothing but the excitation energy of $T+1$ component of the isovector monopole state  in the parent nucleus. 
This is the main component of the monopole state, that is, we can identify it with the main monopole peak that follows the formula  $E_{M_{T+1}}\approx 170 A^{-1/3}$ MeV (cf. Ref.~\cite{auerbach1972} and see Fig.~\ref{fig3}). With this information, Eq.~(\ref{eqa10}) can be simplified as, 
\begin{eqnarray}
  \Delta E_{\rm IAS}&\approx&-\varepsilon^2\frac{N-Z+2}{N-Z}E_{M_{T+1}}\nonumber\\
  &\approx&-170\varepsilon^2\frac{N-Z+2}{N-Z}A^{-1/3} {\rm ~~ MeV} \ .
\label{eqa11}
\end{eqnarray}

\begin{table}[!th]
  \caption{Estimated values of $\Delta E_{\rm IAS}$ from Eq.~(\ref{eqa11}) as caclualated by SAMi. In the second and third columns $\varepsilon^2$ has been calculated using the HF ground state including $V_C$ and $V_C+V_{\rm ISB}$, respectively (see Table \ref{tab4}). In the fourth and fifth columns $\varepsilon^2$ has been calculated using the RPA ground state including $V_C$ and $V_C+V_{\rm ISB}$, respectively (see Table \ref{tab4}). All values are given in MeV.} 
\label{tab5}
\centering
\begin{tabular}{l|cc|cc}
\hline\hline
Nucleus & \multicolumn{2}{c|}{TDA}& \multicolumn{2}{c}{RPA} \\  \hline 
        & $V_{C}$&$V_{C}+V_{\rm ISB}$&$V_{C}$&$V_{C}+V_{\rm ISB}$ \\ \hline  
${}^{48}$Ca & $-0.31$& $-0.29$& $-0.05$& $-0.01$ \\
${}^{90}$Zr & $-0.48$& $-0.44$& $-0.20$& $-0.14$ \\
${}^{132}$Sn& $-0.15$& $-0.14$& $-0.04$& $-0.02$ \\
${}^{208}$Pb& $-0.20$& $-0.19$& $-0.08$& $-0.07$ \\
\hline\hline
\end{tabular}
\end{table}

In the second column of Table \ref{tab5}, we show the energy shifts that estimate the correction to the numerical results calculated from Eq.~(\ref{eq7}) provided in Sec.~\ref{theo:dias}. As it can be seen from the comparison of these values with the third column of Table \ref{tab1}, this correction is about 4\% in ${}^{48}$Ca and about 1\% in ${}^{208}$Pb. The other columns in Table \ref{tab5} give  the energy shifts  produced by the  isospin mixing effects on $E_{\rm IAS}$ due to different $\mathcal{H}_{\rm ISB}$ and different approximations for the ground state wave function. By $V_{\rm ISB}$ we refer to CSB and CIB terms other than Coulomb as included in Ref.~\cite{roca-maza2018}.  

Following the same procedure, one can also estimate the total isospin mixing effect on the energy of the IAS \cite{auerbach1972}. For that, one needs to directly evaluate
\begin{equation}
E_{\rm IAS} = \frac{\langle 0\vert T_+[\mathcal{H},T_-] \vert 0\rangle}{\langle 0\vert T_+T_-\vert 0\rangle} \ ,  
\end{equation}
within the above assumptions. After some algebra, and keeping only terms up to $\varepsilon^2$, one finds
\begin{eqnarray}
  E_{\rm IAS} &=& E_d^{T,~T}+2\varepsilon^2\frac{N-Z+1}{N-Z}\left(E_d^{T+1,~T}-E_d^{T,~T}\right)\nonumber\\
  &&~~~~~~~~-\varepsilon^2\frac{N-Z-2}{N-Z}E_{M_{T+1}}   \nonumber\\
  &=& E_d^{T,~T}-\varepsilon^2\frac{N-Z-2}{N-Z}E_{M_{T+1}},  
\end{eqnarray}
assuming $E_d^{T+1,~T}\approx E_d^{T,~T}$ in the last line. Hence, the total isospin mixing effect on the $E_{\rm IAS}$ is within this simple model
\begin{eqnarray}
  \Delta E_{\rm IAS}^{\rm tot} = -\varepsilon^2\frac{N-Z-2}{N-Z}E_{M_{T+1}} \ ,
\label{ias_tot_im}
\end{eqnarray}
in agreement with Ref. \cite{auerbach1972}. In the case of the IAS and for nuclei with large $N-Z$, the total isospin mixing $\Delta E_{\rm IAS}^{\rm tot}$ is  within the present model very similar to the isospin mixing effect $\Delta E_{\rm IAS}$ needed to correct our numerical results in Sec.~\ref{theo:dias}.

From all these results, we would like to note that isospin mixing effects in  RPA calculations with all ISB terms are expected to be much smaller as shown in columns 4th and 5th in Table \ref{tab5} (cf. $\varepsilon^2$ values in Table \ref{tab4}). Thus, the effect of isospin mixing in the $E_{\rm IAS}$ would be around or below  $1\%$. Note also that isospin mixing effects are larger when only the Coulomb interaction is taken into account (cf. Table \ref{tab4}) simply because other ISB terms display an average attractive nature compensating in part the effect of the repulsive Coulomb potential.

\subsection{Correction to $E_{\rm DIAS}$}
\label{imes_dias}
For the study of the isospin mixing effects on the DIAS energy, 
 we proceed in a similar way to that for the IAS energy. Given the definition of $\vert {\rm DIAS}\rangle$, the energy of the DIAS can be written without approximations as
\begin{equation}
  E_{\rm DIAS} = \frac{\langle 0\vert T_+^2[\mathcal{H},T_-^2] \vert 0\rangle}{\langle 0\vert T_+^2T_-^2\vert 0\rangle} \ .
\label{edias:im}
\end{equation}
In Eq.~(\ref{eq8}) we have, however, assumed $T_+\vert 0\rangle = 0$ and arrived to the expression
\begin{equation}
E_{\rm DIAS} = 2E_{\rm IAS} + \frac{\langle 0\vert[T_+,[T_+,[[\mathcal{H},T_-],T_-]]] \vert 0\rangle}{2(N-Z)(N-Z-1)}.
\label{eq8bis}
\end{equation}
The isospin mixing terms ($\Delta E_{\rm Q.C.}$) left out by our approximation in going from Eq.~(\ref{edias:im}) to Eq.~(\ref{eq8bis})
 can be evaluated as 
\begin{eqnarray}
  \Delta E_{\rm Q.C.} &=& \frac{\langle 0\vert T_+^2[\mathcal{H},T_-^2] \vert 0\rangle}{\langle 0\vert T_+^2T_-^2\vert 0\rangle}\nonumber\\
  &-& 2\frac{\langle 0\vert [T_+,[\mathcal{H},T_-]] \vert 0\rangle}{2T}\nonumber\\
  &-& \frac{\langle 0\vert[T_+,[T_+,[[\mathcal{H},T_-],T_-]]] \vert 0\rangle}{2(N-Z)(N-Z-1)} \ , 
\label{del_qc}
\end{eqnarray}
provided $E_{\rm IAS}$ is calculated as in Sec.~\ref{theo:dias}.  
Adopting the same approximations employed in the previous subsection and keeping terms up to $\varepsilon^2$, we evaluate the three terms in the rhs of the last equation. After some straightforward algebra, we find for the DIAS energy
\begin{eqnarray}
  E_{\rm DIAS} &=& \frac{\langle 0\vert T_+^2[\mathcal{H},T_-^2] \vert 0\rangle}{\langle 0\vert T_+^2T_-^2\vert 0\rangle} \nonumber\\
  &=&E_{2d}^{T,~T} -2\varepsilon^2\frac{N-Z-4}{N-Z-1}E_{M_{T+1}}, \nonumber\\
\label{del_qc_1}  
\end{eqnarray}
where the double displacement energy $E_{2d}^{T,~T}\equiv E_{T,~T-2}-E_{T,~T}$ and it can be approximated as twice the single displacement energy for our purposes here (see Fig.~\ref{fig3} for a schematic representation). That is $E_{2d}^{T,~T}\approx 2 E_{d}^{T,~T}$. Hence, the total isospin mixing effects on the energy of the DIAS can be estimated by the following expression:
\begin{eqnarray}
  \Delta E_{\rm DIAS}^{\rm tot} &=& -2\varepsilon^2 \frac{N-Z-4}{N-Z-1}E_{M_{T+1}} \ . \nonumber\\ 
\end{eqnarray}
The total isospin mixing effect on the energy of the DIAS is compared with that on the energy of the IAS estimated in Eq.~(\ref{ias_tot_im}) as  
\begin{eqnarray}
  \frac{\Delta E_{\rm DIAS}^{\rm tot}}{\Delta E_{\rm IAS}^{\rm tot}} &=& 2\frac{(N-Z)(N-Z-4)}{(N-Z-1)(N-Z-2)}\xrightarrow{T>>1}2 \ .~~~~~~~
\end{eqnarray}
For large isospin imbalance $N-Z>>1$, the correction takes its maximum value which corresponds to twice the correction for the IAS and rapidly drops for small values of $N-Z$, and becomes zero for $N-Z=4$.   Hence, the approximation of assuming a parent state $\vert 0\rangle$ with good isospin is as good (or better) for the DIAS energy as it is for the IAS energy.

The second term in the rhs of Eq.~(\ref{del_qc}) 
 gives
\begin{eqnarray}
\frac{\langle 0\vert [T_+,[\mathcal{H},T_-]] \vert 0\rangle}{2T}&=&E_{d}^{T,~T}+\varepsilon^2\frac{4}{N-Z}E_{M_{T+1}} \ .~~~~
\label{del_qc_2}
\end{eqnarray}
Notice that the factor in $\varepsilon^2$ estimates the isospin mixing effects actually included in our calculations of the IAS energy in Sec.~\ref{theo:dias} via the HF densities employed. That is,  
\begin{eqnarray}
\Delta E_{\rm IAS}^{\rm tot}-\Delta E_{\rm IAS}&=& \frac{4\varepsilon^2}{N-Z}E_{M_{T+1}} \ ,
\end{eqnarray}
and it is clear that its contribution is suppressed by a factor $N-Z$ as compared to the expressions for $\Delta E_{\rm IAS}^{\rm tot}$ and $\Delta E_{\rm IAS}$.

The last term to be evaluated is the Q.C. in Eq.~(\ref{del_qc}),  
\begin{eqnarray}
&&E_{Q.C.}=\frac{\langle 0\vert[T_+,[T_+,[[\mathcal{H},T_-],T_-]]] \vert 0\rangle}{2(N-Z)(N-Z-1)} \nonumber\\
&&= E_{d}^{T,~T-1}-E_{d}^{T,~T} -\varepsilon^2\left(E_{d}^{T,~T-1}-E_{d}^{T,~T}\right)\nonumber\\
&&+3\varepsilon^2\frac{N-Z+1}{N-Z-1}\left(E_{d}^{T+1,~T-1}-E_{d}^{T+1,~T}\right)\nonumber\\
&&-2\varepsilon^2\frac{N-Z+1}{N-Z-1}\frac{N-Z+2}{N-Z}\left(E_{d}^{T+1,~T}-E_{d}^{T+1,~T+1}\right) \ .\nonumber\\
\label{del_qc_3}
\end{eqnarray}
For the Q.C. energy, terms of  $E_{M_{T+1}}$ exactly cancel out and, therefore, only differences on neighbouring single displacement energies appear in the last expression. Those are expected to be small and can be neglected as compared to the isospin mixing effects evaluated in Eqs.~(\ref{del_qc_1}) and (\ref{del_qc_2}). The result in Eq.~(\ref{del_qc_3}) allows us to give a clear physical interpretation to the Q.C. presented in Sec.~\ref{theo:dias}. That is, whenever the isospin mixing effects are neglected, $E_{\rm DIAS}-2E_{IAS}$ tests the actual difference between the displacement energies of the parent $E_{d}^{T,~T}\equiv E_{T,~T-1}-E_{T,~T}$ and daughter $E_{d}^{T,~T-1}\equiv E_{T,~T-2}-E_{T,~T-1}$ nuclei.  

We can now evaluate the isospin mixing effects, $\Delta E_{\rm Q.C.}$,  in Eq. \eqref{del_qc}. Specifically, by using Eqs.~(\ref{del_qc_1}) and (\ref{del_qc_2}) and neglecting 
 Eq.~(\ref{del_qc_3}), we find
\begin{eqnarray}
  \Delta E_{\rm Q.C}&=&-2\varepsilon^2\frac{(N-Z+2)(N-Z-2)}{(N-Z)(N-Z-1)}E_{M_{T+1}} \nonumber\\
&=&2\frac{N-Z-2}{N-Z-1}\Delta E_{\rm IAS}
  \label{del_qc_final}\\
  &&\xrightarrow{N-Z>>1}-2\varepsilon^2E_{M_{T+1}} \ . \label{del_qc_final_app}
\end{eqnarray}

In Table \ref{tab6}, we show the energy shifts that would estimate the correction to the numerical results shown in columns 4th to 6th in Table \ref{tab1} from the quartic commutator given in Eq.~(\ref{eq8}).  Here the isospin mixing $\Delta E_{\rm IAS}$ is neglected in the evaluation of $E_{\rm IAS}$. 
If the  isospin mixing effects are 
  properly  accounted for in $E_{\rm IAS}$ as given in the last column of Table \ref{tab1}, the numerical results for the quartic commutator would need to be corrected by this amount.  Namely, we would need to subtract $2\Delta E_{\rm IAS}$ from Eq.~(\ref{del_qc_final}). In other words,
\begin{eqnarray}
  \Delta \tilde{E}_{\rm Q.C}&=& \Delta E_{\rm Q.C.}-2\Delta E_{\rm IAS} \nonumber\\
  &=& 2\varepsilon^2 \frac{N-Z+2}{(N-Z)(N-Z-1)}E_{M_{T+1}}  \nonumber\\
&=&-\frac{2}{N-Z-1}\Delta E_{\rm IAS} \label{del_qc_final2}\\ 
 &&\xrightarrow{N-Z>>1}\varepsilon^2\frac{E_{M_{T+1}}}{N-Z} \ . \label{del_qc_final2_app}  
\end{eqnarray}

In Table \ref{tab7} we give the energy shifts  \eqref{del_qc_final2}
   provided that $E_{\rm IAS}$ contains all isospin mixing effects. In this case the energy shift due to isospin mixing is positive and smaller as larger is $N-Z$. 
    Our numerical results given in Table \ref{tab1} would be barely corrected by isospin mixing effects.    

\begin{table}[!th]
\caption{ The same as Table \ref{tab5} but for $\Delta E_{\rm Q.C.}$ in Eq.~(\ref{del_qc_final}). Values are given in MeV.} 
\label{tab6}
\centering
\begin{tabular}{l|cc|cc}
\hline\hline
Nucleus & \multicolumn{2}{c|}{TDA}& \multicolumn{2}{c}{RPA} \\  \hline 
        & $V_{C}$&$V_{C}+V_{\rm ISB}$&$V_{C}$&$V_{C}+V_{\rm ISB}$ \\ \hline  
${}^{48}$Ca  & $-0.53$& $-0.50$& $-0.09$& $-0.02$\\
${}^{90}$Zr  & $-0.85$& $-0.78$& $-0.35$& $-0.25$\\
${}^{132}$Sn & $-0.29$& $-0.27$& $-0.08$& $-0.04$\\
${}^{208}$Pb & $-0.39$& $-0.37$& $-0.16$& $-0.14$\\
\hline\hline
\end{tabular}
\end{table}

\begin{table}[!th]
\caption{The same as Table \ref{tab5} but for $\Delta \tilde{E}_{\rm Q.C.}$  in  Eq.~(\ref{del_qc_final2}). Values are given in MeV.} 
\label{tab7}
\centering
\begin{tabular}{l|cc|cc}
\hline\hline
Nucleus & \multicolumn{2}{c|}{TDA}& \multicolumn{2}{c}{RPA} \\  \hline 
        & $V_{C}$&$V_{C}+V_{\rm ISB}$&$V_{C}$&$V_{C}+V_{\rm ISB}$ \\ \hline  
${}^{48}$Ca  & $0.09$&$ 0.08$& $0.01$& $0.003$\\
${}^{90}$Zr  & $0.11$& $0.10$& $0.04$& $0.03$\\
${}^{132}$Sn & $0.01$& $0.01$& $0.003$& $0.001$\\
${}^{208}$Pb & $0.01$& $0.01$& $0.004$& $0.003$\\
\hline\hline
\end{tabular}
\end{table}

As a conclusion, the energy of the $E_{\rm DIAS}$ is little affected by the  isospin mixing effects. However,  the isospin mixing effects are comparable to the quantity $E_{\rm DIAS}-2E_{\rm IAS}$ whenever $E_{\rm IAS}$ is calculated as in Eq.~(\ref{del_qc_2}). On  the contrary, if $E_{\rm IAS}$ and $E_{\rm DIAS}$ contain the  isospin mixing effects, the correction to the quartic commutator results in Sec.~\ref{theo:dias} would be barely changed in most of the studied cases (compare Tables \ref{tab1} and \ref{tab7}). 

\section{$E_{\rm GTR}-E_{\rm IAS}$: commutator evaluation}
\label{app2}

In what follows, we rewrite the interaction (\ref{interac}) in a fully equivalent yet convenient way for the evaluation of the commutators 
 \begin{eqnarray}  
   && V=\sum_i^{A}\kappa_{ls}{\bm l}(i)\cdot{\bm s}(i) \nonumber\\
   &&+\frac{1}{2}\frac{\kappa_{\tau}}{A}\left\{\sum_{i,j}^{A}{\bm \tau}(i)\cdot{\bm \tau}(j)-\sum_{i}^{A}{\bm \tau}(i)\cdot{\bm \tau}(i)\right\} \nonumber \\
   &&+\frac{1}{2}\frac{\kappa_{\sigma}}{A}\left\{\sum_{i,j}^{A}{\bm \sigma}(i)\cdot{\bm \sigma}(j) - \sum_{i}^{A}{\bm \sigma}(i)\cdot{\bm \sigma}(i)\right\}  \nonumber\\
   &&+\frac{1}{2}\frac{\kappa_{\sigma\tau}}{A}\left\{\sum_{i, j}^{A}({\bm \sigma}(i)\cdot{\bm \sigma}(j)) ({\bm \tau}(i)\cdot{\bm \tau}(j) ) \right.\nonumber\\
  &&~~~~~~~~~~~~~~~~-\left.\sum_{i}^{A}({\bm \sigma}(i)\cdot{\bm \sigma}(i)) ({\bm \tau}(i)\cdot{\bm \tau}(i) )\right\}  \ . 
  \nonumber \\
 \end{eqnarray}
Note that due to the properties of the Pauli matrices, 
$\sum_{i}^{A}{\bm \tau}(i)\cdot{\bm \tau}(i) = \sum_{i}^{A}3\hat{\mathds{1}}=3A\hat{\mathds{1}}$, $\sum_{i}^{A}{\bm \sigma}(i)\cdot{\bm \sigma}(i) = \sum_{i}^{A}3\hat{\mathds{1}}=3A\hat{\mathds{1}}$ and $\sum_{i}^{A}({\bm \sigma}(i)\cdot{\bm \sigma}(i)) ({\bm \tau}(i)\cdot{\bm \tau}(i) ) =\sum_{i}^{A}9\hat{\mathds{1}}=9A\hat{\mathds{1}}$ and, therefore, these terms will not contribute to the double or quartic commutators that we evaluate in what follows. 

Firstly we derive the double commutator with the GT operator.  We find,
\begin{widetext}
\begin{eqnarray} 
\left[ O_{+}, \left[\sum_i^{A}{\bm l}(i)\cdot{\bm s}(i) ,  O_{-}\right]\right] &=&-2\sum_i^{A}({\bm l}(i)\cdot{\bm s}(i) -l_z(i) s_z(i)) \label{com-ls}  \ , \\
\left[ O_{+}, \left[\sum_{i, j}^{A}{\bm \sigma}(i)\cdot{\bm \sigma}(j),  O_{-}\right]\right] &=& -4\sum_{i, j}^{A}({\bm \sigma}(i)\cdot{\bm \sigma}(j)- \sigma(i) _z\sigma_z(j)) \nonumber \\
 &&+2\sum_{i, j}^{A}({\bm \sigma}(i)\cdot{\bm \sigma}(j)- \sigma_z(i) \sigma_z(j)) ({\bm \tau}(i)\cdot{\bm \tau}(j)- \tau_z(i) \tau_z(j)) \label{com-s}  \ , \\
\left[ O_{+}, \left[\sum_{i, j}^{A}{\bm \tau}(i)\cdot{\bm \tau}(j),  O_{-}\right]\right] &=&-2\sum_{i, j}^{A}(1- \sigma_z(i) \sigma_z(j)) ({\bm \tau}(i)\cdot{\bm \tau}(j)+\tau_z(i) \tau_z(j)) \label{com-t}  \ , \\
\left[ O_{+}, \left[\sum_{i, j}^{A}({\bm \sigma}(i)\cdot{\bm \sigma}(j))({\bm \tau}(i)\cdot{\bm \tau}(j)),  O_{-}\right]\right] &=& ~~4\sum_{i, j}^{A}({\bm \sigma}(i)\cdot {\bm \sigma}(j)-\sigma_z(i)\sigma_z(j))(1+\tau_z(i) \tau_z(j))\nonumber\\
 &&+2\sum_{i, j}^{A}(1- {\bm \sigma}(i)\cdot {\bm \sigma}(j))({\bm \tau}(i)\cdot{\bm \tau}(j)+\tau_z(i) \tau_z(j)) \label{com-st} \ .
  \end{eqnarray} 
\end{widetext}

For even-even nuclei, there is no contribution from the spin-spin interaction to the previous commutators. 

The average energy is expressed as
 \bea \label{GT-Eapp}
E_{\rm GT}&-&E_{\rm IAS}=\frac{ \langle  0|[ O_{+}, [V,  O_-]]|0 \rangle}{(N-Z)} \nonumber \\
&=&-\frac{4}{3}\frac{\kappa_{ls}}{N-Z}\langle  0|\sum_i^{A}{\bm l}(i)\cdot{\bm s}(i)|0 \rangle \nonumber \\
&+&2( \kappa_ {\sigma \tau}- \kappa_ {\tau}) \frac{N-Z}{A}, 
\eea
since $\langle  0|\sigma\cdot\sigma|0 \rangle=0$ for the spin saturated nuclei.  The expectation value of $\langle 0 |l_zs_z|0 \rangle$ is equal to $\langle 0 |({\bm l}(i)\cdot{\bm s}(i))|0 \rangle/3$ in the spherical nuclei. We stress that in our model it is implicit that all radial matrix elements are equal, and that only the calculation of the direct terms is required for consistency with the assumtion of a separable interaction.

\section{$E_{\rm DGTR}-E_{\rm DIAS}$: commutator evaluation}
\label{app3}
 
Let us now evaluate the quartic commutator in Eq.~(\ref{Hamil32}). After some straightforward algebra, we obtain
\begin{widetext}
\bea 
\langle  0|\left[ O_{+},\left[\left[ O_{+}, \left[\sum_i^{A}{\bm l}(i)\cdot{\bm s}(i),  O_{-}\right]\right],  O_{-}\right]\right]|0 \rangle &=& ~~4\sum_i^{A}({\bm l}(i)\cdot{\bm s}(i) -l_z(i) s_z(i)) \label{qct-ls} \ , \\ 
\langle  0|\left[ O_{+},\left[\left[ O_{+},\left[\sum_{i, j}^{A}{\bm \tau}(i)\cdot{\bm \tau(j)},  O_{-}\right]\right],  O_{-}\right]\right]|0 \rangle &=&~~12\sum_{i, j}^{A} (1- \sigma_z(i) \sigma_z(j))({\bm \tau}(i)\cdot{\bm \tau}(j)+\tau_z(i) \tau_z(j)) \label{qct-t}  \ , \\
\langle  0|\left[ O_{+},\left[\left[ O_{+},\left[\sum_{i, j}^{A}{\bm \sigma}(i)\cdot{\bm \sigma(j)},  O_{-}\right]\right],  O_{-}\right]\right]|0 \rangle &=&~~8\sum_{i, j}^{A}({\bm \sigma}(i)\cdot{\bm \sigma}(j)- \sigma_z(i) \sigma_z(j))(3-\tau_z(i) \tau_z(j)) \nonumber \\
&&-12\sum_{i, j}^{A}({\bm \sigma}(i)\cdot{\bm \sigma}(j)- \sigma_z(i) \sigma_z(j))({\bm \tau}(i)\cdot{\bm \tau}(j)-3\tau_z(i) \tau_z(j)) \ ,\nonumber\\\label{qct-s} \\ 
\langle  0|\left[ O_{+},\left[\left[ O_{+},\left[\sum_{i, j}^{A}{\bm \sigma}(i)\cdot{\bm \sigma(j)}{\bm \tau}(i)\cdot{\bm \tau(j)},  O_{-}\right]\right],  O_{-}\right]\right]|0 \rangle &=&-24\sum_{i, j}^{A}({\bm \sigma}(i)\cdot{\bm \sigma}(j)-\sigma_z(i) \sigma_z(j))(1+\tau_z(i) \tau_z(j)) \nonumber \\
&&+8\sum_{i, j}^{A}({\bm \sigma}(i)\cdot{\bm \sigma}(j)-\sigma_z(i) \sigma_z(j))({\bm \tau}(i)\cdot{\bm \tau}(j)-\tau_z(i) \tau_z(j)) \nonumber \\
&&-4\sum_{i, j}^{A}(1-{\bm \sigma}(i){\bm \sigma}(j))({\bm \tau}(i)\cdot{\bm \tau}(j)+\tau_z(i) \tau_z(j))\nonumber \\
&&-8\sum_{i, j}^{A}(1-\sigma_z(i)\sigma_z(j))({\bm \tau}(i)\cdot{\bm \tau}(j)+\tau_z(i) \tau_z(j)) \label{qct-st} \ .
\eea
\end{widetext}

The energy difference between DGTR and DIAS (\ref{Hamil2}) is now expressed by using the relation in Eq.~(\ref{Hamil32}) as 
\begin{widetext}
\be \label{EGT-EIAS}
E_{\rm DGTR}-E_{\rm DIAS}-\left(1+\frac{N-Z}{N-Z-1}\right)(E_{\rm GT}-E_{\rm IAS})=\frac{4}{3}\frac{\kappa_{ls}\langle  0|\sum_i^{A}{\bm l}(i)\cdot{\bm s}(i)|0 \rangle}{(N-Z)(N-Z-1)}-6( \kappa_ {\sigma \tau}- \kappa_ {\tau}) \frac{1}{A}\frac{N-Z}{N-Z-1}
\ee
\end{widetext}

\section{Determination of $\kappa_{\sigma\tau}-\kappa_\tau$}
\label{app4}

The difference $\kappa_{\sigma\tau}-\kappa_{\tau}$ is estimated from the experimental $E_{\rm GTR}-E_{\rm IAS}$ values in ${}^{48}$Ca \cite{yako2009}, ${}^{90}$Zr \cite{wakasa1997, krasznahorkay2001}, ${}^{112-124}$Sn \cite{pham1995} and ${}^{208}$Pb \cite{akimune1995} as follows. Assuming $V_{ls}=34$ MeV, we find the optimal value for $\kappa_{\sigma \tau}-\kappa_{\tau}$ that reproduce via Eq.(\ref{GT-E}) the experimental results to be $-4$ MeV. We show the results in Table \ref{tab8}.

\begin{table}[!th]
  \caption{Experimental values of $E_{\rm GTR}-E_{\rm IAS}$ (peak energies) in ${}^{48}$Ca \cite{yako2009}, ${}^{90}$Zr \cite{wakasa1997, krasznahorkay2001}, ${}^{112-124}$Sn \cite{pham1995} and ${}^{208}$Pb \cite{akimune1995}, 
together with the predictions from Eq. (\ref{GT-E}), assuming $V_{ls}=34$ MeV and $\kappa_{\sigma\tau}-\kappa_{\tau}=-4$ MeV.}
\label{tab8}
\centering
\begin{tabular}{lccclcc}
\hline\hline
Nucleus & \multicolumn{2}{c}{$E_{\rm GTR}-E_{\rm IAS}$}&~~~~&Nucleus & \multicolumn{2}{c}{$E_{\rm GTR}-E_{\rm IAS}$}\\
&Exp&Eq.(\ref{GT-E})& ~~~~ & &Exp&Eq.(\ref{GT-E}) \\
& [MeV] & [MeV]&~~~~& &[MeV] & [MeV]\\
\hline
${}^{48}$Ca  &3.3& 3.82&& ${}^{118}$Sn &1.3& 0.87\\
${}^{90}$Zr  &3.9& 3.63&& ${}^{120}$Sn &1.2& 0.53\\
${}^{112}$Sn &2.8& 3.04&& ${}^{122}$Sn &1.0& 0.65\\
${}^{114}$Sn &2.1& 2.60&& ${}^{124}$Sn &1.0&0.73 \\
${}^{116}$Sn &1.7& 1.64&& ${}^{208}$Pb &0.4& 0.42\\
\hline\hline
\end{tabular}
\end{table}

\bibliography{bibliography.bib}

\end{document}